\documentclass[]{spie}  

\usepackage{amsmath,amsfonts,amssymb}
\usepackage{graphicx}
\usepackage[colorlinks=true, allcolors=blue]{hyperref}

\title{Should we trade off higher-level mathematics for abstraction to improve student understanding of quantum mechanics?}

\author[a]{James K. Freericks}
\author[a]{Leanne Doughty}
\affil[a]{Department of Physics, Georgetown University, Washington, DC 20057, USA}

\authorinfo{Further author information: (Send correspondence to J.K.F.)\\J.K.F.: E-mail: james.freericks@georgetown.edu}

\begin{document} 
\maketitle

\begin{abstract}
Undergraduate quantum mechanics focuses on teaching through a wavefunction approach in the position-space representation. This leads to a differential equation perspective for teaching the material. However, we know that abstract representation-independent approaches often work better with students, by comparing student reactions to learning the series solution of the harmonic oscillator versus the abstract operator method. Because one can teach all of the solvable quantum problems using a similar abstract method, it brings up the question, which is likely to lead to a better student understanding? In work at Georgetown University and with edX, we have been teaching a class focused on an operator-forward viewpoint, which we like to call \textit{operator mechanics}. It teaches quantum mechanics in a representation-independent fashion and allows for most of the math to be algebraic, rather than based on differential equations. It relies on four fundamental operator identities---(i) the Leibniz rule for commutators; (ii) the Hadamard lemma; (iii) the Baker-Campbell-Hausdorff formula; and (iv) the exponential disentangling identity. These identities allow one to solve eigenvalues, eigenstates and wavefunctions for all analytically solvable problems (including some not often included in undergraduate curricula, such as the Morse potential or the P\"oschl-Teller potential). It also allows for more advanced concepts relevant for quantum sensing, such as squeezed states, to be introduced in a simpler format than is conventionally done. In this paper, we illustrate the three approaches of matrix mechanics, wave mechanics, and operator mechanics, we show how one organizes a class in this new format, we summarize the experiences we have had with teaching quantum mechanics in this fashion and we describe how it allows us to focus the quantum curriculum on more modern 21st century topics appropriate for the second quantum revolution.
\end{abstract}


\section{INTRODUCTION}
\label{sec:intro}  

We are in the midst of a second quantum revolution---one in which we can detect, control, and manipulate individual quanta---electrons, atoms, and photons, to name a few. This scientific advance is likely to usher in a new era in advanced technology---an era where technologies employ these abilities to detect, control, and manipulate individual quanta in a fashion that creates novel perhaps even life-changing technologies (as the global positioning system has, which is based on atomic clocks). The science is also new. It is called quantum information science, with its three pillars of quantum computing, quantum communication, and quantum sensing.  In order to reap the benefits of the second quantum revolution, we need to have a workforce that is substantially more quantum literate than we currently have.\cite{workforce} We need workers who will be quantum aware, quantum proficient, and quantum expert. Teaching the principles of quantum information science, especially for the quantum aware and quantum proficient workforce, in a manner that uses lower mathematical prerequisites makes the material more accessible and the field more inclusive.  Quantum computing and quantum communications are already taught in nontraditional ways. Many instructors employ the idea that quantum mechanics is simply a novel way in which to calculate probabilities from complex-valued probability amplitudes that satisfy rules of superposition. In addition, quantum gates, and quantum channels can be described using finite-sized matrix representations and the principles of linear algebra. However, in the area of quantum sensing one needs to understand advanced concepts such as atomic structure, photon-matter coupling, squeezed states, decoherence, and a realistic description of measurement in actual experiments. This is generally believed to require a treatment that uses the conventional differential equation approach. This statement is wrong. In this paper, we discuss strategies that trade-off more advanced mathematics with abstraction to allow the quantum sensing pillar of quantum information science to also be taught in a more accessible fashion and with less prerequisites.

The old Bohr-Sommerfeld theory for quantum mechanics operated under an assumption that quantum mechanics resulted from restrictions on classical mechanics. The Bohr-Sommerfeld quantization rules related phase-integrals along classical trajectories as constraints that are quantized in units of $\hbar$. It worked remarkably well for hydrogen---it properly determined the energy levels, the relativistic corrections and all fine-structure effects. But it lacked a proper theoretical grounding. In 1925, Heisenberg discovered \textit{matrix mechanics},\cite{heisenberg} which replaced continuous classical functions with discrete (but infinite) matrices that satisfy the same equations of motion as the Hamiltonian equations of motion for classical mechanics do. But, because the matrix starts with the ground state, one can create ladder operators that allow one to compute the allowed energy eigenvalues, without determining the energy eigenstates. Indeed, in the original matrix mechanics approach, there wasn't even the concept of a a quantum state---all effects were inferred from the matrices themselves. 
Both the harmonic oscillator and the angular momentum problem can be solved in the matrix-mechanics way, and the conventional approaches in most textbooks closely follow this Heisenberg, Born, and Jordan methodology.\cite{heisenberg,born-jordan,drei-mann-arbeit} In current quantum instruction, we work with ladder operators, but rather than thinking of them as infinite-dimensional discrete matrices, we instead think of them as abstract operators. It is spin (and in some case orbital angular momentum) problems that are usually treated with finite-dimensional matrix representations, using a modern matrix mechanics approach, that emphasizes linear algebra methods to solve problems. In this fashion, most students already learn some matrix mechanics ideas in their quantum education.

In 1926, Schr\"odinger discovered \textit{wave mechanics},\cite{schroedinger-wave-mechanics} which introduced a wavefunction in the position-space representation, and used the Schr\"odinger equation to analyze quantum phenomena. This introduces the standard differential-equation approach to quantum mechanics, which is the mainstay of most quantum instruction. In the original Schr\"odinger work, he solved the differential equation using the Laplace method, which expresses the solution as a contour integral in the complex plane. This methodology was rapidly replaced by the Fr\"obenius method, which used a generalized power-series ansatz to solve the differential equations. Most instructors know that students struggle with learning this approach, and most do not master it well enough to solve any new problems that are different from the standard problems appearing in nearly all quantum textbooks.

One might think that this is it. These are the two ways to solve for quantum mechanics. Indeed, there are many more methods, including the Wigner function approach, path integrals, and so forth. A recent publication outlined nine different ways to ``do'' quantum mechanics.\cite{styer9}~ But, there is one more method that is not mentioned elsewhere. It is what we call \textit{operator mechanics} and it was invented the same time that matrix mechanics and wave mechanics was invented. The first operator mechanics solution was Pauli's solution for hydrogen.\cite{pauli} While this is often described as a matrix mechanics solution, a quick perusal of the paper shows that he calculates everything by manipulating operators and their commutation relations. He determines the two separate SU(2) symmetries that combine to make up the SO(4) symmetry of the Coulomb problem. The ladder operators of this Lie algebra then determine the energy spectra of hydrogen. But, again, it does not determine the energy eigenstates. Dirac also worked with similar ideas when he described q-numbers and c-numbers,\cite{q-numbers-c-numbers} indicating the notion that one can solve quantum problems by working with the algebra of the q-numbers (operators). Modern operator mechanics uses operators to compute all of the same properties that are ordinarily taught in a quantum mechanics classroom. In 1940, Schr\"odinger invented the factorization method,\cite{factorization} which generalizes the abstract method for the harmonic oscillator to all exactly solvable quantum mechanics problems. These operator methods can even be used to find the wavefunctions\cite{freericks-rushka,squeezing,cartesian-hydrogen,rodrigues} by employing the translation operator and other operator identities. This approach requires an ability to work with abstraction, but does not require as much advanced mathematics as conventional approaches.

We show how these three different approaches inter-relate when we solve the harmonic oscillator three different ways below.

In traditional instruction, one has to teach the students significant amounts of advanced mathematics, including series solutions to differential equations, Fourier transforms, Dirac delta functions, and how to work with differential operators. The time spent doing this takes away from the time that can be spent on teaching physical ideas related to experiments and how to measure properties of quantum objects. Many of us have experienced this with our students who often lament how the quantum class was more of a math class than a physics class. When one teaches the class using operator mechanics, it does not require anywhere near as much time spent learning new math, which allows much more physics to be covered. This allows us to modernize the curriculum to teach modern experiments from the second quantum revolution, such as Bell experiments, delayed choice experiments, interaction-free measurements, single quanta detection, the Hong-Ou-Mandel effect and much more. In the classes we teach, we also discuss how one measures the momentum of a single quanta, which is a topic that is not covered in most textbooks---instead many instructors teach incorrect statements such as ``it is impossible to measure position and momentum at the same time.'' We describe below how one actually measures momentum, and what is wrong with the above statement---all without violating the Heisenberg uncertainty principle.\cite{momentum} 

As we consider the options available to broaden quantum instruction for 21st century applications, we have to grapple with three options: (i) make students interested in this area take the standard physics curriculum and all of the standard mathematics and physics prerequisites; (ii) find a way to simplify the math load by solving all problems using linear algebra methods and requiring only linear algebra as the math prerequisite; or (iii) reformulate how quantum mechanics is taught so that the required math is the math commonly learned in high school. While many educators lean towards scenarios (i) or (ii), we focus on scenario (iii). To do this requires one to put operators first and to introduce and work with abstract quantum states. Hence, one needs to teach students to work more abstractly in order to lower the mathematics prerequisites. There is a tension then between whether students without significant math prerequisites are able to handle abstraction or whether one needs to increase the required prerequisites. In other words, do students have the cognitive load to enable working more abstractly? We argue in this work that students can handle the abstraction and that option (iii) is a viable one. It is likely to be the approach that will be most inclusive as well (because it sharply reduces the prerequisites).

And what about option (ii)? Is it possible to teach linear algebra without the calculus sequence? This should certainly be true for most of linear algebra (one would need to avoid teaching about how polynomials can be a vector space, for example). We feel there are two serious challenges with the linear-algebra approach though. If we relegate all of the calculations to manipulating matrices (probably numerically) then we expect it will be difficult for students to develop conceptual understanding of the material. Everything they do would involve constructing a matrix, manipulating it, and then extracting the desired results. The operator-based approach is different from this. While operators can be represented by matrices and one can work with the representations using linear algebra, one can also just work with them abstractly and this is the way we do it when we teach quantum mechanics in this fashion. We do find students appear to develop some conceptual ideas associated with these operator manipulations.

In 2020, the National Science Foundation held a workshop to develop the key concepts needed by quantum information science learners.\cite{key-concepts} The participants developed a sequence of nine key concepts, which cover qubits, superposition, entanglement, measurement, and more. It provides a topical framework for developing quantum curriculum for the 21st century. Our quantum class used these key concepts as the initial framework for determining the high-level learning goals, which we discuss in more detail below.

\section{OPERATOR MECHANICS VERSUS WAVE AND MATRIX MECHANICS}

Since so many textbooks treat quantum mechanics in identical ways, we wanted to take a moment to describe how each of the three approaches---matrix mechanics, wave mechanics, and operator mechanics---treat the simple harmonic oscillator in one dimension, which is one of the most important quantum problems.

Few people are now familiar with the original matrix mechanics of Heisenberg, Born and Jordan. We take the opportunity to show how the original reasoning works, so one can understand the approach. In modern quantum instruction, matrix mechanics usually refers to problems that have a finite number of degrees of freedom, and can be treated using a matrix representation for the operators. The simplest example is in covering spin (or more generally two-level systems) by employing the Pauli spin matrices. While the original approach shares many similarities with this, it is different. The description here follows closely the Born and Jordan paper\cite{born-jordan} from 1925.

The original idea of Heisenberg is that the dynamical degrees of freedom $q$ and $p$ are described by infinite-dimensional matrices. The Hamiltonian is constructed from the identical functional form as in classical mechanics, but with the classical variables replaced by the matrix ones. For a given dynamical system, the matrix elements of position and momentum vary harmonically with the energy differences (this was one of Heisenberg's postulates), so we have
\begin{equation}
    \textbf{q}_{mn}=q_{mn}e^{i\omega_{mn}t}~~\text{and}~~\textbf{p}_{mn}=p_{mn}e^{i\omega_{mn}t},
\end{equation}
where the bold symbol denotes an (infinite-dimensional discrete) matrix and $\omega_{mn}=\tfrac{E_m-E_n}{\hbar}$. We also use the classical equations of motion. So $\dot{\textbf{q}}=\tfrac{1}{m}\textbf{p}$ and $\dot{\textbf{p}}=-\tfrac{\partial \textbf{U}(\textbf{q})}{\partial{\textbf{q}}}=-m\omega^2\textbf{q}$ for the simple harmonic oscillator, because $\textbf{H}=\tfrac{\textbf{p}^2}{2m}+\tfrac{m\omega^2\textbf{q}^2}{2}$, with $\omega$ the frequency of the harmonic oscillator. The position and momentum operators are real-valued in classical mechanics, which leads to the postulate that their matrices must be Hermitian, so $q_{nm}=q_{mn}^*$ and $\omega_{nm}=-\omega_{mn}$; note that this last result is consistent with the Rydberg-Ritz combination principle.

These matrices also satisfy the canonical commutation relation, which Born and Jordan called the \textit{quantum condition}:
\begin{equation}
    [\textbf{q},\textbf{p}]=\textbf{q}\textbf{p}-\textbf{p}\textbf{q}=i\hbar\textbf{I},
\end{equation}
with $\textbf{I}$ the (infinite-dimensional and discrete) identity matrix. Using the relation from the equations of motion, that $p_{mn}=i\omega_{mn}q_{mn}$, and the Hermiticity, we find the canonical commutation relation becomes
\begin{equation}
    im\sum_l(\omega_{ln}-\omega_{ml})q_{ml}q_{ln}=i\hbar \delta_{mn}.
\end{equation}
Setting $m=n$, and using the Hermiticity, we find that
\begin{equation}
    \sum_l\omega_{ml}|q_{ml}|^2=-\frac{\hbar}{2m}.
\end{equation}

The classical equation of motion (re-expressed using the substitution of matrices for the quantum treatment) is $(\omega_{mn}^2-\omega^2)q_{mn}=0$. This says that whenever $q_{mn}\ne 0$, then $|\omega_{mn}|=\omega$. Now, we use another postulate, that the energy eigenvalues are nondegenerate, so we have at most two nonzero values for $q_{mn}$: one with $\omega_{mm'}=\omega$ and one with $\omega_{mm"}=-\omega$ (note that not both solutions need exist). There must be at least one solution always, otherwise the canonical commutation relation will not hold.

Our next step is to compute the diagonal elements of $\textbf{H}$, which is postulated to be the allowed energy levels of the system (you can think of this as reading off the eigenvalues of the Hamiltonian by looking at the diagonal elements of \textbf{H} when expressed in an energy eigenbasis because Heisenberg's matrices are required by their construction to be representations in the energy eigenbasis). We find that
\begin{equation}
    H_{mm}=\frac{m}{2}\left [(-\omega_{mm'}\omega_{m'm}+\omega^2)|q_{mm'}|^2+(-\omega_{mm"}\omega_{m"m}+\omega^2)|q_{mm"}|^2\right ]=m\omega^2(|q_{mm'}|^2+|q_{mm"}|^2).
\end{equation}
Now, we order our indices, such that $m=0$ is the lowest energy, which must be positive, because the harmonic oscillator Hamiltonian is the sum of two positive terms. For $m=0$, only $q_{01}$ and $q_{10}$ are nonzero. The canonical commutation relation tells us that $|q_{01}|^2=-\tfrac{\hbar}{2m\omega_{01}}$ or $E_0=H_{00}=\tfrac{1}{2}\hbar\omega$, because $\omega_{01}=-\omega$. This is the ground-state energy. The condition that $\omega_{mm'}=\pm\omega$, then tells us that the general energy levels satisfy $E_n=\hbar\omega(n+\tfrac{1}{2})$, because excited states have higher energies, which is the standard result. There is nothing more to determine, except matrix elements for a specific problem, because this approach does not use quantum states or wavefunctions explicitly in its approach.

Note that we did not solve this using a factorization of the Hamiltonian followed by stepping up and down the spectrum with the ladder operators. That approach turns out to easily arise from this matrix mechanics formalism and it appeared in Born and Jordan's first quantum mechanics textbook,\cite{born-jordan-textbook} which was written in 1930 and came out a number of months \textit{before} Dirac's first edition of \textit{The Principles of Quantum Mechanics}\cite{dirac-principals-first-ed} (which was also published in 1930). But, it is interesting to note that the standard method that appears in most textbooks is actually the old matrix-mechanics approach of Heisenberg, Born, and Jordan, and not the invention of Dirac, as many contemporaries would  now say.

Next, we move on to the differential equation approach, which is laid out the same way it is done in modern textbooks in Schr\"odinger's paper detailing its solution.~\cite{factorization} We start from the Schr\"odinger energy eigenvalue equation for the time-independent system in position space, given by
\begin{equation}
    -\frac{\hbar^2}{2m}\frac{d^2\psi_n(q)}{dq^2}+\frac{1}{2}m\omega^2q^2\psi_n(q)=E_n\psi_n(q),
\end{equation}
where $\psi_n(q)$ is the $n$th energy eigenstate wavefunction. Here, $E_n$ is the energy eigenvalue for the $n$th energy eigenstate of the simple harmonic oscillator. In Schr\"odinger's original treatment,\cite{schroedinger-2} he solves the problem by showing the differential equation and just stating the solution. But, we will show how the standard approach works, which employs the Fr\"obenius method of generalized series solutions. We first examine the asymptotic behavior, where $|q|\to\infty$, and the energy eigenvalue can be neglected in the differential equation. The functional behavior of the wavefunction can be easily worked out to be of the form $\psi_n(q)=f_n(q)e^{-\tfrac{m\omega}{2\hbar}q^2}$, where we must have that $f_n(q)$ does not diverge too rapidly to make the wavefunction not normalizable. The differential equation for $f_n$ becomes
\begin{equation}
    -\frac{\hbar^2}{2m}\frac{d^2 f_n(q)}{dq^2}+\hbar\omega q \frac{d f_n(q)}{dq}+(\tfrac{1}{2}\hbar\omega-E_n)f_n(q)=0.
\end{equation}
The equation is solved by a series solution ansatz, given by
\begin{equation}
    f_n(q)=\sum_{m=0}^\infty c_mq^m.
\end{equation}
Then we find that the coefficients satisfy
\begin{equation}
    c_{m+2}=\frac{2m\omega}{\hbar}\frac{\hbar\omega(m+\tfrac{1}{2})-E_n}{(m+2)(m+1)}c_m.
\end{equation}
Doing an asymptotic analysis of the power series shows that $f_n(q)$ grows too quickly unless the coefficients terminate after a finite number of them. This is accomplished by having $E_n=\hbar\omega(n+\tfrac{1}{2})$. In this case, one finds that the polynomials $f_n(q)$ are the Hermite polynomials. Normalization, at the end produces a fully normalized wavefunction. We won't go through these details any further, because they appear in nearly all textbooks and should be familiar to all.

Operator mechanics has its origins in Dirac's early paper\cite{q-numbers-c-numbers} on $c$ numbers and $q$-numbers, where we now would call $q$ numbers abstract operators. But, the first real operator mechanics paper is Pauli's original solution of the eigenvalues of hydrogen.\cite{pauli}. That paper is often described as a matrix mechanics calculation, but a careful examination of the paper finds that all of the calculations of the commutators are done abstractly starting from $[\hat{q},\hat{p}]=i\hbar$; in this work, we use hats to denote operators. Indeed, this is all that is used to calculate the commutation relations that establish the SO(4) symmetry. Then using the ladder operators, one can find the energy eigenvalues. Pauli did not realize he had discovered the SO(4) symmetry. That had to wait until Fock described it later.~\cite{fock-hydrogen}

The modern formulation of operator mechanics comes from Schr\"odinger's work during World War Two.\cite{factorization,factorization2,factorization3} In that work, he developed what is called the \textit{factorization method}. This is a method by which one factorizes the original Hamiltonian in the form $\hat{H}_0=\hat{A}_0^\dagger\hat{A}_0^{\phantom{\dagger}}+E_0$. Because this is a positive semidefinite operator, the lowest energy state satisfies $\hat{A}_0|\psi_0\rangle=0$ (called the \textit{subsidiary condition}), yielding a ground-state energy of $E_0$. Next, we form the first auxiliary Hamiltonian via $\hat{H}_1^{aux}=\hat{A}_0^{\phantom{\dagger}}\hat{A}_0^\dagger+E_0$, which produces a \textit{different} Hamiltonian from the original one; note how we interchanged the order of the ladder operators for the first auxiliary Hamiltonian. We then factorize the auxiliary Hamiltonian in terms of new raising and lowering operators $\hat{H}_1^{aux}=\hat{A}_1^\dagger\hat{A}_1^{\phantom{\dagger}}+E_1$. This gives a new subsidiary condition to find its ground state $\hat{A}_1^{\phantom{\dagger}}|\phi_1^{aux}\rangle=0$, with an energy given by $E_1$; we use $\phi$ to denote the auxiliary Hamiltonian eigenstates. The \textit{factorization chain} is constructed by repeating this procedure, starting next with $\hat{H}_1^{aux}$, and finding the next auxiliary Hamiltonian, energy, and ground state. The procedure produces two forms for each auxiliary Hamiltonian. They can be written as 
\begin{equation}
    \hat{H}_j^{aux}=\hat{A}_{j-1}^{\phantom{\dagger}}\hat{A}_{j-1}^\dagger+E_{j-1}=\hat{A}_j^\dagger\hat{A}_j^{\phantom{\dagger}}+E_j.\label{eq:auxiliaryH}
\end{equation}
What is amazing is the energy  $E_1$ is also the energy of the first excited state of $\hat{H}_0$ (and $E_j$ is the energy of the $j$th excited state of $\hat{H}_0$). This comes from the fact that the construction of the auxiliary Hamiltonians gives us the \textit{intertwining relation}, which follows from the two forms that we can use to represent each auxiliary Hamiltonian in terms of the ladder operators in Eq.~(\ref{eq:auxiliaryH}). The intertwining relation says that
\begin{equation}
    \hat{H}_j^{aux}\hat{A}_j^\dagger=\hat{A}_j^\dagger\hat{H}_{j+1}^{aux}
\end{equation}
This can also be used to show that $|\psi_1\rangle\propto \hat{A}_0^\dagger|\phi_1^{aux}\rangle$, because
\begin{equation}
    \hat{H}_0|\psi_1\rangle=\hat{H}_0\hat{A}_0^\dagger|\phi_1^{aux}\rangle=\hat{A}_0^\dagger\hat{H}_1^{aux}|\phi_1^{aux}\rangle=E_1\hat{A}_0^\dagger|\phi_1^{aux}\rangle=E_1|\psi_1\rangle.
\end{equation}
Similarly, the $n$th excited state has energy given by $E_n$ and 
\begin{equation}
    |\psi_n\rangle=\frac{1}{\sqrt{(E_n-E_{n-1})(E_n-E_{n-2})\cdots(E_n-E_0)}}\hat{A}_0^\dagger\hat{A}_1^\dagger\cdots\hat{A}_{n-1}^\dagger|\phi_n^{aux}\rangle,
\end{equation}
where the normalization factor can be easily worked out as well from the intertwining relation and the auxiliary ground state energies because $\hat{A}_n^{\phantom{\dagger}}\hat{A}_n^\dagger=\hat{H}_{n+1}-E_n$.

You can see that this is an abstract approach, but it allows us to determine energy eigenvalues without wavefunctions (or differential equations). The procedure follows from relatively simple factorizations, because the solvable problems have \textit{shape-invariant potentials}. When one has a shape invariant potential, the factorizations always have the same functional form, just different coefficients for the individual terms, so it is a simple exercise to determine all of the required factorizations needed to find the auxiliary Hamiltonians. 

What about wavefunctions? They can also be found algebraically by using an operator generalization of the Rodrigues formulas.\cite{cartesian-hydrogen,rodrigues} The strategy starts by rewriting the ladder operators as a similarity transformation of the momentum operator. Then, the subsidiary condition is used to find a state built off of the $n$th auxiliary ground state that is annihilated by the momentum operator. Using such a state, the $n$th energy eigenstate can be converted into an $n$-fold iterated commutator expression acting on the auxiliary Hamiltonian ground state (this is the analog of the Rodrigues formula). Finally, one uses recurrence relations to find the appropriate polynomial multiplied by another function to give us the wavefunction.  

We next describe how one uses this approach for the simple harmonic oscillator. While this approach is treated in nearly all textbooks, it often is described as a special case applicable only to the harmonic oscillator and often the operators are simply presented without showing how to determine them via the factorization method. Here, we will present the arguments along the lines of how Schr\"odinger originally did this, which uses a different normalization of the ladder operators (so some aspects may look unfamiliar to you). We also show an additional \textit{singular} factorization that is rarely discussed (yes, the factorization need not be unique). It arises because the potential is even. It has additional surprises as well.

The factorization method was developed in the 1940s and then was reinvigorated in the 1980s when Witten used it as an example of supersymmetry.\cite{witten-susy} Supersymmetric quantum mechanics and the factorization method are closely related to each other. We will not discuss supersymmetry further here, but we will use the naming conventions from supersymmetric quantum mechanics.

We start with the ladder operators, which are Hermitian conjugates of each other. The lowering operator is expressed as
\begin{equation}
    \hat{A}=\frac{1}{\sqrt{2m}}\big (\hat{p}-i\hbar kW(k'\hat{q})\big ),
\end{equation}
where $W$ is the superpotential and $k$ and $k'$ are wavenumbers with dimensions of inverse length. Calculating the product of the ladder operators yields
\begin{equation}
\hat{A}^\dagger\hat{A}=\frac{\hat{p}^2}{2m}-\frac{i\hbar k}{2m}[\hat{p},W(k'\hat{q})]+\frac{\hbar^2 k^2}{2m}W^2(k'\hat{q}).
\end{equation}
The procedure for factorizing the Hamiltonian then requires us to find a superpotential such that
\begin{equation}
    V(\hat{q})-E=-\frac{i\hbar k}{2m}[\hat{p},W(k'\hat{q})]+\frac{\hbar^2 k^2}{2m}W^2(k'\hat{q}).\label{eq:potential}
\end{equation}
For solvable problems, the superpotential always takes the form of a simple expression involving the sum over one to three simple terms. The additional factorizations employ the exact same form, which is why the solvable problems are said to have shape-invariant potentials. This is what makes the solution of these factorizations feasible. 

Note that if we already know the ground-state wavefunction $\psi_0(q)$, then we can directly verify that $\hbar k W(k'\hat{q})=-\tfrac{\psi_0^\prime(\hat{q})}{\psi_0(\hat{q})}$, which follows from the well-known transformation from the Schr\"odinger equation to the Riccati equation. Of course, if we are solving the problem, we cannot know the wavefunction \textit{a priori}, hence the solution proceeds by constructing the superpotential directly. Because there can be multiple factorizations that produce the same Hamiltonian (that is, the factorization method is not unique), we simply state some of the requirements for a valid superpotential (which are developed fully in the class). First, a nonsingular factorization (with a state that exists that satisfies the subsidiary condition) is always unique. By nonsingular, we mean that the superpotential does not diverge inside the domain of the wavefunction (it can diverge at the boundary points). If the superpotential is singular, it describes a wavefunction with nodes at every point where it is singular. Second, the superpotential is real for any problem that does not have linear terms in momentum in the Hamiltonian. And third, the sign of the superpotential must be positive for $q\to\infty$ and negative for $q\to -\infty$ in order for the wavefunction to be normalizable. This can be seen to be a requirement by looking at the relationship between the superpotential and the logarithmic derivative of the wavefunction.

We are now ready to compute the factorization for the simple harmonic oscillator. A clear choice is to pick a linear function, so we can try $W(k'\hat{q})=k'\hat{q}$. Substituting this into Eq.~(\ref{eq:potential}), gives us
\begin{equation}
\frac{1}{2}m\omega^2\hat{q}^2-E=-\frac{\hbar^2 kk'}{2m}+\frac{\hbar^2k^2k'^2}{2m}\hat{q}^2,
\end{equation}
which requires us to have $\hbar kk'=\pm m\omega$. Using the sign requirement for the superpotential (which guarantees there is a normalizable state that satisfies the subsidiary condition), then yields
\begin{equation}
    \hat{A}_0^{\phantom{\dagger}}=\frac{1}{\sqrt{2m}}\big(\hat{p}-im\omega\hat{q}\big),
\end{equation}
and $E_0=\tfrac{1}{2}\hbar\omega$, which are the standard results (except for a rescaling of the ladder operators versus the contemporary scaling). Note that this superpotential is not singular over the entire domain for the wavefunction $-\infty<q<\infty$, while it does diverge at the boundary points. 

We might have thought this is it, but there is another factorization one can find. The way to think of this is to try a superpotential of the form $\tfrac{a}{\hat{q}}+b\hat{q}$. Then the commutator with momentum will yield a term that is an inverse square power in $\hat{q}$ and a constant, while the square of the superpotential yields an inverse squared term, a constant and a squared term. If we can adjust the constants so that the inverse square term has a vanishing coefficient, we will have another factorization of the simple harmonic oscillator. Indeed, such a procedure does work. Let's sort out the details. We first need to show how we calculate the commutator of momentum with $\tfrac{1}{\hat{q}}$ without using differentiation. We obtain it instead from the Leibniz rule. Noting that $[\hat{p},1]=0$, or that the momentum commutes with a number, we find
\begin{equation}
    0=[\hat{p},1]=\left [\hat{p},\tfrac{\hat{q}}{\hat{q}}\right ]=\hat{q}\left [\hat{p},\tfrac{1}{\hat{q}}\right ]+[\hat{p},\hat{q}]\tfrac{1}{\hat{q}}=\hat{q}\left [\hat{p},\tfrac{1}{\hat{q}}\right]-\tfrac{i\hbar}{\hat{q}},
\end{equation}
which can be rearranged to yield $\left [\hat{p},\tfrac{1}{\hat{q}}\right ]=\tfrac{i\hbar}{\hat{q}^2}$. This same strategy is used to compute all more complex commutators. It turns out that one can actually work out all of the rules of derivatives solely by working with commutators, because commutators satisfy the Leibniz rule and that is all that is needed to obtain the rules of differentiation,\cite{robinson}~ but we do not discuss this further here.

Using this result, we set the superpotential to be
\begin{equation}
 kW(k'\hat{q})=\frac{a}{\hat{q}}+b\hat{q},
\end{equation}
with $a\ne 0$ and $b>0$. Then we find that
\begin{equation}
    \frac{1}{2}m\omega^2\hat{q}^2-E=\frac{\hbar^2}{2m}\left (\frac{a}{\hat{q}^2}-b\right )+\frac{\hbar^2}{2m}\left (\frac{a^2}{\hat{q}^2}+2ab+b^2\hat{q}^2\right ).
\end{equation}
To have a new factorization ($a\ne 0$) and to satisfy the sign requirement for the superpotential when $|q|$ is large, we must pick $a=-1$ and $b=\tfrac{m\omega}{\hbar}$. Hence, we have
\begin{equation}
    \hat{A}_0^{'\phantom{\dagger}}=\frac{1}{\sqrt{2m}}\left (\hat{p}+\frac{i\hbar}{\hat{q}}-im\omega\hat{q}\right ),
\end{equation}
and $E'=\tfrac{3}{2}\hbar\omega$. We use primes for the ladder operators and the energies for the singular factorization to distinguish them from the nonsingular factorization. This is the first excited-state energy, not the ground state. But that is to be expected, because the factorization is singular, meaning the lowest energy state that we can find has a node at $q=0$ (because the superpotential diverges there), which cannot be the ground state because the ground state is nodeless. This is an important result for the additional (singular) factorizations of a Hamiltonian. They provide results for some, but not all of the energy eigenstates. We will discover below that this factorization only finds all of the eigenstates with odd parity. 

The next step is to construct the auxiliary Hamiltonians. To do this we compute the product of the ladder operators in the opposite order. This means we take the original Hamiltonian and add $[\hat{A}_0^{\phantom{\dagger}},\hat{A}_0^\dagger]=\tfrac{i\hbar k}{m}[\hat{p},W(k'\hat{q})]$ to find the first auxiliary Hamiltonian.  For the nonsingular factorization, it is equal to $\hat{H}_1^{aux}=\tfrac{\hat{p}^2}{2m}+\tfrac{1}{2}m\omega^2\hat{q}^2+\hbar\omega$, which shifts the original Hamiltonian upwards by the constant $\hbar\omega$. Now, to factorize this Hamiltonian, we find we use \textit{exactly the same} ladder operators as we used for the original Hamiltonian, only now the constant term has increased by $\hbar\omega$ to $\tfrac{3}{2}\hbar\omega$. Because the ladder operators are the same, the process continues for the next auxiliary Hamiltonian, and so on. Hence, we have $\hat{H}_n^{aux}=\tfrac{\hat{p}^2}{2m}+\tfrac{1}{2}m\omega^2\hat{q}^2+n\hbar\omega$ and $E_n=\hbar\omega\left (n+\tfrac{1}{2}\right )$. This is the only problem that has such simple auxiliary Hamiltonians.

The singular case proceeds similarly, but in this case the auxiliary Hamiltonians change more than just being shifted upwards in energy. Computing $[\hat{A}_0^{'\phantom{\dagger}},\hat{A}_0^{'\dagger}]=\tfrac{\hbar^2}{m}\left (\tfrac{1}{\hat{q}^2}+\tfrac{m\omega}{\hbar}\right)$, then yields
\begin{equation}
    \hat{H}_1^{aux~'}=\frac{\hat{p}^2}{2m}+\frac{\hbar^2}{m\hat{q}^2}+\frac{1}{2}m\omega^2\hat{q}^2+\hbar\omega,
\end{equation}
which is a potential with a singularity at the origin. Experts will recognize this Hamiltonian as the $l=1$ radial Hamiltonian (with $\hat{q}\to\hat{r}$ and $\hat{p}\to \hat{p}_r$) for the three-d isotropic harmonic oscillator. Now, we need to factorize the Hamiltonian again using the same ansatz as before, but now with a new $a$ and $b$. We find in this case we must choose $a=-2$ and $b=\tfrac{m\omega}{\hbar}$, which yields $E_1'=\tfrac{5}{2}\hbar\omega$. Note, that one might have thought there is ambiguity about whether we pick $a=-2$ or $a=1$, since both yield the same coefficient. But we must choose the negative value, because the energy must be positive, so there is no ambiguity. Now, we can repeat this approach and we find the general result is
\begin{equation}
    \hat{A}_n=\frac{1}{\sqrt{2m}}\left (\hat{p}+\frac{i\hbar(n+1)}{\hat{q}}-im\omega\hat{q}\right ),~~E_n'=\hbar\omega\left (2n+\tfrac{3}{2}\right )~~\text{and}~~\hat{H}_n^{aux~'}=\frac{\hat{p}^2}{2m}+\frac{\hbar^2 n(n+1)}{2m\hat{q}^2}+\frac{1}{2}m\omega^2\hat{q}^2+n\hbar\omega.
\end{equation}
The energy increases by $2\hbar\omega$ with each step in the factorization chain, because this result only determines the odd solutions of the original Hamiltonian.

Now, we compute the energy eigenstates following the factorization method recipe. We will use the normalization factor to ensure we construct normalized states. For the nonsingular factorization, we find that we obtain
\begin{equation}
    |\psi_n\rangle=\frac{1}{\sqrt{(\hbar\omega)^n}}\frac{1}{\sqrt{n!}}\left (\hat{A}^\dagger\right )^n|\psi_0\rangle,
\end{equation}
for the state with $E_n=\hbar\omega\left (n+\tfrac{1}{2}\right )$, which should look quite familiar. There is an extra factor in the normalization, because the ladder operators do not have the standard normalization. The $(\hbar\omega)^nn!$ term comes from the product $(E_n-E_0)(E_n-E_1)\cdots(E_n-E_{n-1})$ in the denominator of the normalization factor. Finally, we just use one raising operator, because we found for this factorization chain that all raising operators are the same. Now, the situation is a bit different for the singular factorization. Here, we find that
\begin{equation}
    |\psi_n'\rangle=\frac{1}{\sqrt{(\hbar\omega})^n}\frac{1}{\sqrt{(2n)!!}}\hat{A}_0^{'\dagger}\hat{A}_1^{'\dagger}\cdots\hat{A}_{n-1}^{'\dagger}|\psi_0'\rangle,
\end{equation}
which are the energy eigenstates with $E_n'=\hbar\left (2n+\tfrac{3}{2}\right )$. It might not seem like this is correct, but these states correspond to the states formed with the nonsingular potential for odd index, where $(n)_{nonsing}=2(n)_{singular}+1$, up to a possible global phase. This can be verified directly by constructing the states, but we do not do so here. 

You might be surprised that you have never seen this singular factorization before. It was always there, waiting to be discovered---it simply required a clear understanding of how the factorization method works to find it!

Now, we move on to computing wavefunctions. We first want to remind how wavefunctions are found when one works in the representation-independent fashion. The wavefunction is the overlap of the quantum state (which we will take to be an energy eigenstate), with the position eigenstate bra. Hence, we have $\psi_n(q)=\langle q|\psi_n\rangle$. The position eigenstate satisfies $\hat{q}|q\rangle=q|q\rangle$, where $q$ is a number that tells us the location of the position eigenstate. It can be constructed by acting the position translation operator onto the position eigenstate at the origin, via
\begin{equation}
|q\rangle=e^{-\tfrac{i}{\hbar}q\hat{p}}|0_q\rangle,
\end{equation}
where $|0_q\rangle$ is the position eigenstate at the origin (which is annihilated by the position operator, namely $\hat{q}|0_q\rangle=0$). To see this, we simply compute
\begin{equation}
    \hat{q}|q\rangle=\hat{q}e^{-\tfrac{i}{\hbar}q\hat{p}}|0_q\rangle=e^{-\tfrac{i}{\hbar}q\hat{p}}e^{\tfrac{i}{\hbar}q\hat{p}}\hat{q}e^{-\tfrac{i}{\hbar}q\hat{p}}|0_q\rangle=e^{-\tfrac{i}{\hbar}q\hat{p}}(\hat{q}+q)|0_q\rangle=qe^{-\tfrac{i}{\hbar}q\hat{p}}|0_q\rangle=q|q\rangle,
\end{equation}
because $\hat{q}|0_q\rangle=0$. Note that we used the Hadamard lemma in the middle of the derivation to compute the shift of the position operator by $q$ after the similarity transformation with the translation operator.

For the wavefunctions, we will concentrate on the nonsingular factorization, as the calculation is a bit long. We begin with finding the similarity transformation of the momentum operator that yields the raising operator. Using the Hadamard lemma (or expanding the Gaussian in a power series and evaluating the commutator term by term), we find that
\begin{equation}
    \hat{A}^\dagger=\frac{1}{\sqrt{2m}}(\hat{p}+im\omega\hat{q})=\frac{1}{\sqrt{2m}}e^{\tfrac{m\omega}{2\hbar}\hat{q}^2}\hat{p}e^{-\tfrac{m\omega}{2\hbar}\hat{q}^2}.
\end{equation}
The energy eigenstate can then be written as
\begin{equation}
    |\psi_n\rangle=\frac{1}{\sqrt{(2\hbar m\omega)^n}}\frac{1}{\sqrt{n!}}e^{\tfrac{m\omega}{2\hbar}\hat{q}^2}\hat{p}^ne^{-\tfrac{m\omega}{2\hbar}\hat{q}^2}|\psi_0\rangle,
\end{equation}
because the interior Gaussian factors combine to yield 1. The next step is to find the state annihilated by the momentum operator. We find this state by applying the similarity transformation to the subsidiary condition, which yields the result
\begin{equation}
    \hat{p}\underbrace{e^{\tfrac{m\omega}{2\hbar}\hat{q}^2}|\psi_0\rangle}_{\text{state}}=0.
\end{equation}
Because the momentum operator annihilates this state, and because we can write $e^{-\tfrac{m\omega}{2\hbar}\hat{q}^2}=e^{-\tfrac{m\omega}{\hbar}\hat{q}^2}e^{\tfrac{m\omega}{2\hbar}\hat{q}^2}$, we can rewrite the first momentum operator acting to the left as a commutator, which gives us
\begin{equation}
|\psi_n\rangle=\frac{1}{\sqrt{(2\hbar m\omega)^n}}\frac{1}{\sqrt{n!}}e^{\tfrac{m\omega}{2\hbar}\hat{q}^2}\hat{p}^{n-1}\left [\hat{p},e^{-\tfrac{m\omega}{\hbar}\hat{q}^2}\right ]e^{\tfrac{m\omega}{2\hbar}\hat{q}^2}|\psi_0\rangle.
\end{equation}
We can repeat this operation of introducing commutators $n-1$ more times, and we can then combine all exponential operators to the left, since a function of the position operator commutes with the nested momentum operators to obtain
\begin{equation}
    |\psi_n\rangle=\frac{1}{\sqrt{(2\hbar m\omega)^n}}\frac{1}{\sqrt{n!}}e^{\tfrac{m\omega}{\hbar}\hat{q}^2}\left [\hat{p},\left [\hat{p},\cdots\left [\hat{p},e^{-\tfrac{m\omega}{\hbar}\hat{q}^2}\right ]\cdots\right]\right]_n|\psi_0\rangle.
\end{equation}
This is the operator generalization of the Rodrigues formula (the $n$ subscript indicates there are $n$ nested commutators). The set of nested commutators will give us Hermite polynomials, which we now show.

We define the operator-valued Hermite polynomial via
\begin{align}
    H_n\left(\hat{q}\sqrt{\frac{m\omega}{\hbar}}\right)|\psi_0\rangle&=(-i)^n\sqrt{\frac{2^n}{(\hbar\omega)^n}}(\hat{A}^\dagger)^n|\psi_0\rangle=(-i)^n\sqrt{2^nn!}|\psi_n\rangle\nonumber\\
    &=\frac{(-i)^n}{\sqrt{(\hbar m\omega)^n}}e^{\tfrac{m\omega}{\hbar}\hat{q}^2}\left [\hat{p},\left [\hat{p},\cdots\left [\hat{p},e^{-\tfrac{m\omega}{\hbar}\hat{q}^2}\right ]\cdots\right]\right]_n|\psi_0\rangle.
\end{align}
Let's evaluate this for $n=0$ and $n=1$. For $n=0$, we find $H_0\left(\hat{q}\sqrt{\tfrac{m\omega}{\hbar}}\right)=1$ and for $n=1$, we have $H_1\left(\hat{q}\sqrt{\tfrac{m\omega}{\hbar}}\right)=2\hat{q}\sqrt{\tfrac{m\omega}{\hbar}}$. These are the first two Hermite polynomials. To establish the remainder, we need to work out the recurrence relation of the Hermite polynomials.

To start, we rewrite the definition of the Hermite polynomial operator as 
\begin{align}
    H_n\left(\hat{q}\sqrt{\frac{m\omega}{\hbar}}\right )&=-i\sqrt{\frac{2}{\hbar\omega}}\hat{A}^\dagger(-i)^{n-1}\sqrt{\frac{2^{n-1}}{(\hbar\omega)^{n-1}}}(\hat{A}^\dagger)^{n-1}|\psi_0\rangle\nonumber\\
    &=-i\sqrt{\frac{2}{\hbar\omega}} \left (\hat{A}+i\sqrt{2m}\omega\hat{q}\right )(-i)^{n-1}\sqrt{\frac{2^{n-1}}{(\hbar\omega)^{n-1}}}(\hat{A}^\dagger)^{n-1}|\psi_0\rangle\nonumber\\
    &=2\hat{q}\sqrt{\frac{m\omega}{\hbar}}H_{n-1}\left (\hat{q}\sqrt{\frac{m\omega}{\hbar}}\right )|\psi_0\rangle+(-i)^n\sqrt{\frac{2^n}{(\hbar\omega)^n}}\left [\hat{A},(\hat{A}^\dagger)^{n-1}\right ]|\psi_0\rangle,
\end{align}
where we used the identity that $\hat{A}^\dagger=\hat{A}+i\sqrt{2m}\omega\hat{q}$. The commutator on the last line is formed due to the subsidiary condition, because $\hat{A}|\psi_0\rangle=0$. Using the fact that  $[\hat{A},\hat{A}^\dagger]=\hbar\omega$,
one can use induction to immediately show that
\begin{equation}
    \left [\hat{A},(\hat{A}^\dagger)^{n}\right ]=n\hbar\omega(\hat{A}^\dagger)^{n-1}. 
\end{equation}
Then, the Hermite polynomial recurrence relation becomes
\begin{equation}
    H_n\left (\hat{q}\sqrt{\frac{m\omega}{\hbar}}\right )|\psi_0\rangle=2\hat{q}\sqrt{\frac{m\omega}{\hbar}}H_{n-1}\left (\hat{q}\sqrt{\frac{m\omega}{\hbar}}\right )|\psi_0\rangle-2(n-1)H_{n-2}\left (\hat{q}\sqrt{\frac{m\omega}{\hbar}}\right )|\psi_0\rangle.
\end{equation}
This is the standard recurrence relation for the Hermite polynomials using the physicist normalization.

Putting these results together, we find that
\begin{equation}
    \psi_n(q)=\langle q|\psi_n\rangle=\frac{i^n}{\sqrt{2^nn!}}\langle q|H_n\left (\hat{q}\sqrt{\frac{m\omega}{\hbar}}\right )|\psi_0\rangle.
\end{equation}
Now, we can act the Hermite polynomial operator to the left against the position eigenstate to obtain
\begin{equation}
    \psi_n(q)=\frac{i^n}{\sqrt{2^nn!}}H_n\left (q\sqrt{\frac{m\omega}{\hbar}}\right )\psi_0(q).
\end{equation}
What remains is the computation of the ground-state wavefunction. But, before doing that, we will drop the $i^n$ factor, since it is a global phase, and we would rather work with a real-valued wavefunction.

The ground-state wavefunction is found by using the fact that the momentum operator annihilates a Gaussian operator acting on the ground state. Hence,
\begin{align}
    \psi_0(q)=\langle q|\psi_0\rangle&=\langle q|e^{-\tfrac{m\omega}{2\hbar}\hat{q}^2}e^{\tfrac{m\omega}{2\hbar}\hat{q}^2}|\psi_0\rangle=e^{-\tfrac{m\omega}{2\hbar}q^2}\langle q|e^{\tfrac{m\omega}{2\hbar}\hat{q}^2}|0\rangle=e^{-\tfrac{m\omega}{2\hbar}q^2}\langle 0_q|e^{\tfrac{i}{\hbar}q\hat{p}}\underbrace{e^{\tfrac{m\omega}{2\hbar}\hat{q}^2}|\psi_0\rangle}_{\hat{p}~\text{annihilates this state}}\nonumber\\
    &=e^{-\tfrac{m\omega}{2\hbar}q^2}\langle 0_q|e^{-\tfrac{m\omega}{2\hbar}\hat{q}^2}|\psi_0\rangle=e^{-\tfrac{m\omega}{2\hbar}q^2}\langle 0_q|0\rangle.
\end{align}
We operated $\hat{q}$ to the left in the last line, where it is replaced by $0$. All that remains is to compute the normalization constant, which is done by a simple Gaussian integral, which yields $\langle 0_q|0\rangle=\left (\tfrac{m\omega}{\pi\hbar}\right )^{\tfrac{1}{4}}$. Putting this all together, we arrive at our normalized wavefunction
\begin{equation}
    \psi_n(q)=\left (\tfrac{m\omega}{\pi\hbar}\right )^{\tfrac{1}{4}}\frac{1}{\sqrt{2^nn!}}H_n\left (q\sqrt{\frac{m\omega}{\hbar}}\right )e^{-\tfrac{m\omega}{2\hbar}q^2},
\end{equation}
which is the standard result.
Note that there are no differential equations used in the derivation, and the only calculus needed is for normalization of the ground state. The normalization constant can be told to students who have not taken calculus, so this aspect should not limit students who have not yet studied calculus from participating.

Working within the operator-mechanics paradigm requires us to focus on how to develop a comfort level for abstraction with the students. As we describe below, this is done by first focusing on conceptual ideas with a simple concrete formalism, and then slowly developing the abstract formal aspects by re-solving the same conceptual problems using the new formal language. This, coupled with extensive practice allows students to master the ability to work more abstractly. It builds upon skills they have already developed in algebra and geometry classes in high school. We discuss more details of this implementation below.

As we have seen, the ``operator mechanics'' approach is reminiscent of the \textit{algebraic} method for determining the energy eigenfunctions and eigenvalues of the simple harmonic oscillator (and of angular-momentum), which appears in many quantum textbooks.
Two modern undergraduate textbooks develop the operator mechanics method fully, one by Green~\cite{green} and one by Ohanian~\cite{ohanian}.
This strategy has recently been extended to obtain spherical harmonics~\cite{weitzman}, simple harmonic oscillator wavefunctions~\cite{freericks-rushka}, squeezed-state wavefunctions~\cite{squeezing}, and hydrogen wavefunctions~\cite{mathews,cartesian-hydrogen} algebraically.
\textit{Every analytically solvable problem treated in conventional quantum mechanics courses can also be treated from a purely algebraic standpoint.} 
Even time evolution can be developed algebraically by using its unitary and semigroup requirements as captured with the Trotter product formula~\cite{trotter}, which also clarifies what a time-ordered product is. 
We have empirically found that working with operators to perform calculations provides a more ``physical feel of moving operators'' to the algebraic methods used by the students, which we have observed helps them master the technical skills and apply this knowledge more broadly to other problems.
Operator mechanics does require tools to work with. All standard quantum calculations can be performed by using four different operator identities (and generalizations of them). These are (i) the Leibniz rule for the commutator of a product of operators $[\hat{A},\hat{B}\hat{C}]=\hat{B}[\hat{A},\hat{C}]+[\hat{A},\hat{B}]\hat{C}$; (ii) the Hadamard lemma $e^{\hat{A}}\hat{B}e^{-\hat{A}}=\hat{B}+[\hat{A},\hat{B}]+\tfrac{1}{2}[\hat{A},[\hat{A},\hat{B}]]+\tfrac{1}{3!}[\hat{A},[\hat{A},[\hat{A},\hat{B}]]]+\cdots$; (iii) the Baker-Campbell-Hausdorff formula, which relates $e^{\hat{A}}e^{\hat{B}}=e^{\hat{A}+\hat{B}+\tfrac{1}{2}[\hat{A},\hat{B}]+\cdots}$; and (iv) the exponential disentangling identity, which allows the exponential of Lie-algebra elements to be re-expressed in terms of products of exponentials of Lie-algebra elements and can be thought of as a special case of the so-called KHK-theorem and Cartan decomposition of a Lie group. These operator identities can be taught using no high-level math, and how to use them is learned by the students as they see examples and work through problems; in particular, no group theory is needed to work with this material.

\section{PHYSICS EDUCATION KNOWLEDGE BASE FOR TEACHING QUANTUM MECHANICS}

Examining student understanding and difficulties in quantum mechanics is an ongoing area of interest in physics education research (PER)~\cite{singh1}. 
This type of work has identified many areas of difficulty with quantum mechanics (\textit{e.~g.}~tunnelling~\cite{wittmann2005}, time dependence~\cite{emigh2015}, measurement and expectation values~\cite{singh2,passante2015}) and has led to a variety of teaching innovations~\cite{zhu2012a,zhu2012b}.
The PER group at the University of Washington has expanded their tutorial approach, successful at the introductory level, to upper division quantum mechanics~\cite{emigh2018}. In fact a complete class of lectures, preflights, tutorials, homeworks, and exams is available in a flexible, modular format from recent work by another group.\cite{pollock}~
Many have chosen to develop visualizations to address student difficulties and improve students conceptual understanding (\textit{e.~g.}~QuILTs~\cite{singh3,singh4,singh5}, PhET~\cite{mckagan2008}, and QuVis~\cite{kohnle2012}). 
Another area of focus has been on testing and comparing the spin-first paradigm and the wave-function-first paradigm with the spin-first approach leading to higher student performance on research-based concept assessments~\cite{sadaghiani2016}. 
The two assessments that have been used to test these innovations and approaches are the Quantum Mechanics Concept Assessment~\cite{sadaghiani2015} and a survey focused on student understanding of quantum mechanics in one spatial dimension~\cite{singh6}. 
In our work, we have adopted visualization strategies but we are not able to use these previously developed instruments that relate to measurement. 
The main issue is that those instruments are based on a technical approach that uses the Copenhagen interpretation and assumes the measurements are nondestructive. Very few of the experimental methods to actually measure individual quanta operate in this fashion.
Instead, we teach that nearly all measurements of single quanta are destructive since students who will work in quantum sensing need to understand how the real-world experiments are actually performed.  
This is why our materials do not discuss measurement in the conventional way, except to explain how it is dangerous to take an ontic view of the wavefunction, because if one does, then one has to reconcile the collapse of a physical wavefunction, say for a photon coming from a star millions of light years a way, with how the knowledge that the photon has been detected at one point is signalled to all of the other points of the wavefunction, including those millions of light years away, in an instant. 
Furthermore, there are no experiments that can measure the position of a particle precisely (causing a collapse to a delta function), so the assessment instruments that use (i) state preparation (typically in an energy eigenstate or a superposition of energy eigenstates for a particle in a box), (ii) measurement of the position exactly, and (iii) then ask how the wavefunction further evolves, are very far from real experiments. 
In our classes, we instead focus on real measurement techniques, such as time-of-flight, which actually measures the position (by a destructive particle detector, or other means) and momentum (via timing the travel time and knowing the distance to the source) in each shot. 
This then allows one to carefully explain how the uncertainty principle works in measurement, applying to many shots, not to each individual shot (as is incorrectly stated in many quantum textbooks). 
We also describe how single-particle detectors work and why they are often destructive (in order to amplify the signal). 
Finally, we discuss interaction-free measurements, which are nondestructive when successful, but can destroy the particle when they fail. 
This is the modern information needed by learners moving into quantum sensing, and it is for this reason that we cannot employ standard assessment instruments to gauge student learning in our course. 

Student difficulties in applying mathematical tools (especially calculus tools) to understand and solve problems in upper-division physics courses has been widely documented \cite{caballero2015}. 
A recent study by Tu \textit{et al.}~\cite{tu2020} identifies the difficulties students have working with partial differential equations in the context of an energy eigenfunction problem in two spatial dimensions and a time evolution problem in one-dimension. 
Even simple arithmetic involving complex numbers has been found to be difficult for undergraduates \cite{smith2015,smith2019}. 
On the other hand, Wawro \textit{et al.}~\cite{wawro2020} have published promising research showing student abilities to critique and understand the purposes of linear algebra and Dirac notations in quantum mechanics and demonstrated students flexibility in reasoning about linear algebra and quantum concepts; this indicates that students should be cognitively able to work with abstract operator algebra.






Many of the identified conceptual difficulties in quantum courses originate from the abstract nature of the material \cite{Bouchee2022}, including interpreting counter-intuitive phenomena and understanding the limitations of language to express quantum concepts.
In our class, we must have the students develop the ability to work with abstraction. We do this with a multipronged approach. 
A mathematics education theory for teaching abstraction \cite{AbstractModel} suggests that you should start with models and representations that introduce the underlying structure of an idea and then work across models and representations to demonstrate extensions to new components and applications. In many topics that we introduce (such as the Mach-Zehnder interferometer, as an example), we first introduce the material using a concrete model that evaluates quantum probabilities in a rules-based methodology, with the quantum rules simply given to the students. Even within such an approach, students gain intuition about quantum phenomena, especially interference. Next, we develop a formal approach employing Dirac notation and operators, and revisit the exact same material using the new formalism. Here, we are able to derive the concrete rules used earlier within a coherent, yet abstract, mathematical framework. Finally, we revisit the material again using a second-quantization approach, where the photon is modeled by the real and imaginary parts of the field excitations of the electric field, which are quantized in a large box. By the time students have reached this final topic, their ability to work with abstraction has grown significantly from where it was before the class began. 

Another approach we use to scaffold the development of abstraction is to assign writing assignments to the students. These short essays (one to two paragraphs) encourage the students to express abstract ideas via analogies, or via more concrete models, which break down a complex system into smaller concrete parts. Students often invest significant thought into how to organize and express these ideas in their writing.

\begin{figure}
    \centering
    \includegraphics[width=3.0in]{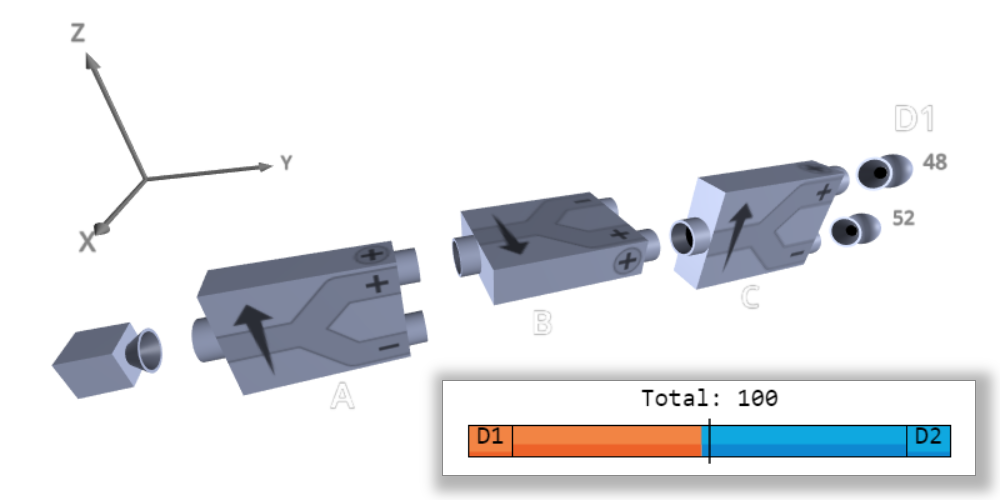}
    \hspace{2em}
    \includegraphics[width=2.5in]{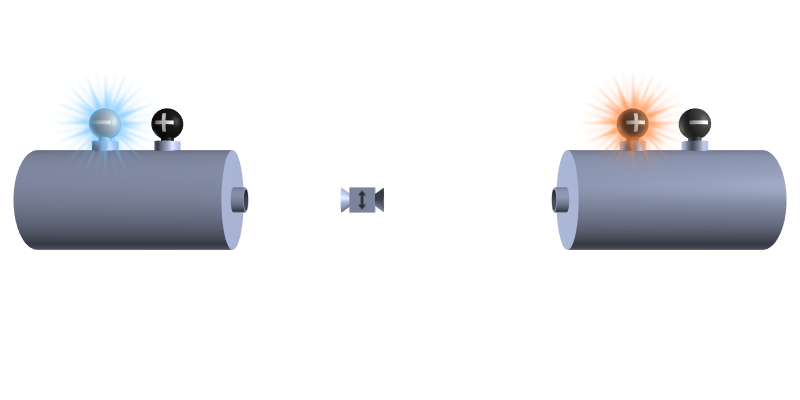}\\
    \vspace{2em}
    \includegraphics[width=2.5in]{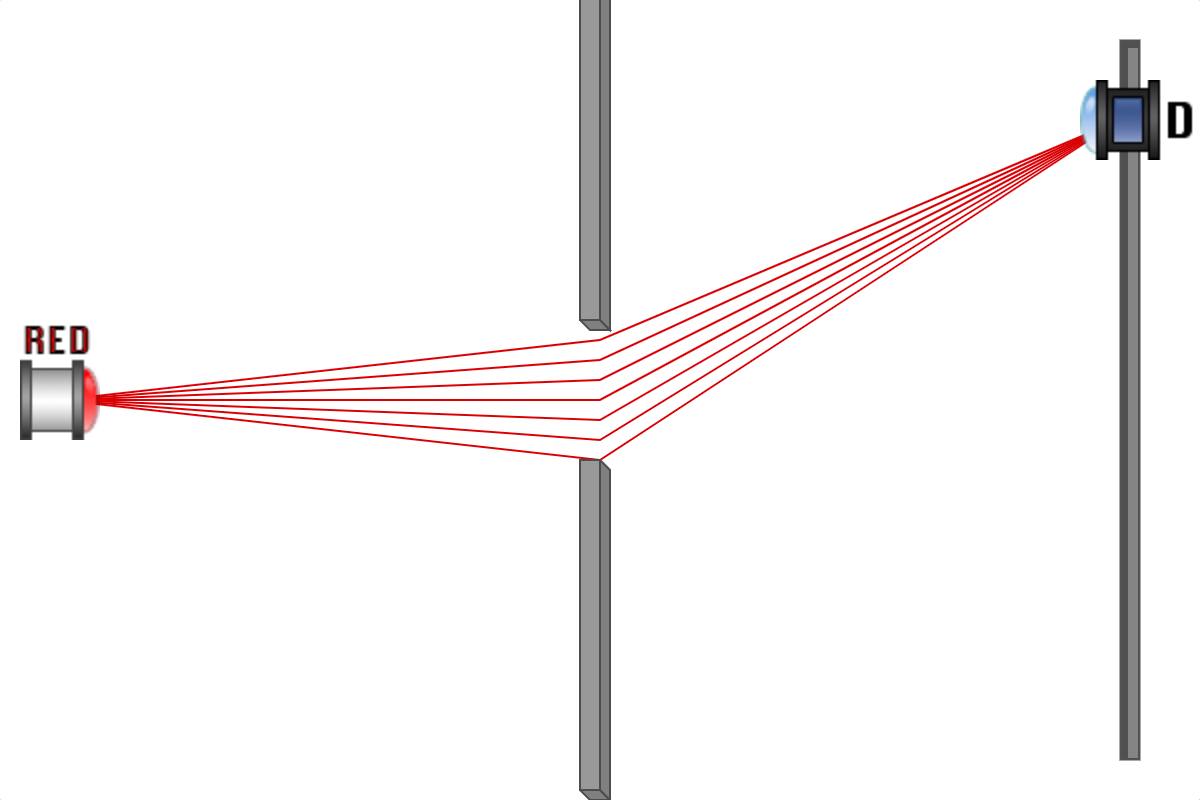}
    \includegraphics[width=0.5in]{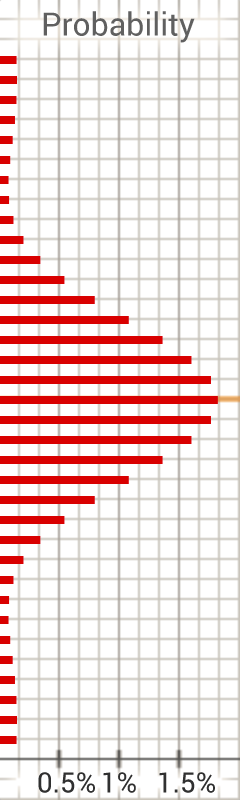}
    \hspace{3em}
    \includegraphics[width=2.5in]{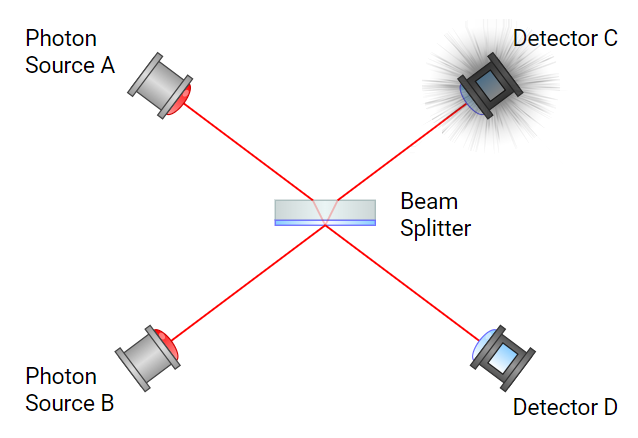}
    \caption{Images from different computer animations in the quantum-mechanics class. Top left: A nested three-Stern-Gerlach analyzer experiment. Top right: A Bell experiment. Bottom left: A single-slit diffraction experiment (illustrating the \textit{wave} properties of single-slit diffraction). Bottom right: The Hong-Ou-Mandel experiment.}
    \label{fig:graphics}
\end{figure}

An innovative feature of the course is the use of embedded javascript simulations, animations, and tutorials.~\cite{physics_teacher,edx-mooc} These elements provide unique visualization and interactive capabilities. An example of some static images from the animations are shown in Fig.~\ref{fig:graphics}.  These types of digital materials assist in abstraction by supporting students' meaning-making in a number of ways \cite{Bouchee2022}: visualizing unobservable quantum phenomena, contrasting classical and quantum behaviour and highlighting surprising and unexpected results.

Finally, by having all assignments (except for exams) immediately graded, with multiple attempts at full credit, we push students from process one thinking toward process two thinking (in the dual-process model for how students reason on tasks \cite{JBST2013,Speirs2021}). 
This ``sparking''  to question their default model should also aid in students refinement of their mental models of abstract concepts. 



\section{ADVICE FOR IMPLEMENTATIONS}

We teach these classes remotely on the edX platform\cite{edx-mooc,physics-teacher} (as a massive open on-line class) or in person as a flipped class. In both configurations, students watch videos before class time. During the class, learners work on problems and short readings or animations that discuss material related to the lecture. Students also have problem sets and examinations, which are also offered in the on-line format. The only difference between the fully on-line class and the flipped class is that instructors are available in person to help students in the flipped class. We have also made available voluntary readings related to the course materials, which are a draft of a book on operator mechanics being written by one of the authors.

The edX framework allows for sophisticated student inputs to answer questions. They can respond to traditional problems such as multiple choice and select all that apply. But, they can also input numerical answers (graded within a tolerance) and symbolic answers too. It is the latter aspect that allows us to replicate the sophisticated nature of traditional quantum classes. But, because the work is computer graded, it also allows students to get immediate feedback, correct incorrect answers and try again (each question has multiple attempts allowed). The platform also allows students to enter essays that answer questions requiring a synthesis of reasoning. Students grade these questions on their own following an instructor-provided rubric. After students have answered questions successfully (or have exhausted all available attempts), solutions are also instantly available. This allows students to immediately redress any errors in logic, which we believe will help them learn the material more effectively. In fact, we believe this approach helps students develop a growth mindset with respect to this material, which should also enhance their ability to learn it.\cite{growth-mindset}

We initiated the course development with a backwards design, starting from high-level learning goals.
The learning goals for the quantum course were based upon the nine core principles for QIS~\cite{key-concepts}
that were identified by an NSF Workshop. 
These core principles form the jumping off point for the quantum course design. 
In addition, we needed to focus on the mathematics required to develop the formalism to ensure accessibility. 
Traditional instruction is often a challenge for students in quantum mechanics, because the level of mathematics required in conventional quantum courses can be overwhelming to a typical student. 
By treating the quantum material in a nontraditional way, we have been able to reduce the core mathematical requirements to a bare minimum and focus more on the science and engineering ideas. 
 
A standard language for describing the knowledge and skills we want students to gain is the revised 2001 Bloom's taxonomy~\cite{bloom}, which describes six different cognitive processes in working with knowledge (with increasing sophistication): (i) Remembering; (ii) Understanding; (iii) Applying; (iv) Analyzing; (v) Evaluating; and (vi) Creating. 
This taxonomy also classifies the types of knowledge as (i) Factual, (ii) Conceptual, (iii) Procedural, and (iv) Metacognitive. 
Given the hierarchical/spiral nature of the curriculum, we were also be informed by the Shavelson and Huang Framework of Cognitive Outcomes~\cite{shavelson} associated with knowledge that is (i) Declarative; (ii) Procedural; (iii) Schematic; and (iv) Strategic. 
This framework recognizes that the cognitive demand of a procedural task is not always greater than for a declarative task. 
Our descriptions of learning goals use both content-based descriptors and student-action/cognitive demand descriptors. 

The overriding goal for the quantum mechanics course is to prepare students to thoughtfully answer the following four questions: 
(i) How do we interrogate the world of the ultrasmall and learn properties of objects too small to be seen by the eye; 
(ii) How do quantum properties emerge from the canonical commutation relation and the commutation relations of spin; 
(iii) What role do superposition and entanglement play in determining how quantum objects behave and how can these properties be employed in applications; and 
(iv) How do we develop a mathematical formalism that encompasses these ideas and provides us with predictive capabilities. 
These broad questions primarily encompass the first five cognitive processes in working with knowledge. 
Students should also emerge with the associated technical skills that allow them to work within the mathematical formalism of quantum mechanics and to be able to apply these skills within new contexts.

The philosophy that underpins this course design process is based on providing the widest \textit{accessibility} to learners with different backgrounds. 
Informed by education research (for example, see the University of Washington's work on tutorials in introductory physics\cite{mcdermott1992research}), we know that students perform far better in science, technology, engineering, and mathematics (STEM) fields when they develop a strong conceptual understanding of the ideas \textit{before} they represent those ideas within a mathematical framework. This approach is counter to the popular textbook by Griffiths~\cite{griffiths}, which instead argues for immediate exposure to the full quantum formalism, and is closer to the ``spins-first'' paradigm~\cite{townshend,mcintyre}, but emphasizes conceptual ideas much more than what is currently present in that regimen.
Our course is also designed using a tiered approach that revisits topics as they are developed more deeply.

The quantum mechanics course has four main parts: (i) conceptual ideas, (ii) technical developments for working with operators; (iii) applications to experiment; and (iv) applications to sensing. 
It begins with conceptual ideas associated with spins and light, which allow us to discuss complex phenomena, such as Bell experiments, nondemolition experiments, and photon bunching. 
Then we develop the formal methods needed to work with operators, including the four fundamental operator identities. 
Next, we apply the formal developments to quantum problems, employing the Schr\"odinger factorization method and relating to many quantum experiments (discovery of the deuteron, the proton charge radius, radio astronomy, nuclear magnetic resonance, molecular rotational spectroscopy, and many more). 
We end by describing how single photons are detected, what a squeezed vacuum is and how Laser Interferometer Gravitational-Wave Observatory (LIGO) can measure distances small enough that it can detect gravitational waves. 
This course develops the first five cognitive processes from Bloom's taxonomy. 

The high-level learning goals for the quantum mechanics course are:
\begin{enumerate}
\item Be able to describe how the principle of superposition underlies wave-particle duality and to use the quantum superposition of states to analyze properties of spin (Stern-Gerlach experiments) and light (one-, two-, and multiple slits and the Mach-Zehnder interferometer).
\item Distinguish between events, alternative ways an event occurs, reversible tagging, and irreversible measurement; be able to use the mathematical formalism of quantum mechanics to describe these different phenomena and how they allow for delayed choice experiments.
\item Describe how entanglement requires quantum mechanics to violate local realism and how experiments verify that this occurs.
\item Explain how an interaction-free experiment allows for quantum seeing in the dark and the details of how such an experiment is carried out.
\item Know the four fundamental operator identities of quantum mechanics (Leibniz rule, Hadamard lemma, Baker-Campbell-Hausdorff formula, and exponential disentangling identity), in all of their variants, and be able to efficiently use them in quantum calculations with operators.
\item Use the Schr\"odinger factorization method to solve bound-state energy eigenvalue problems (determining the energies algebraically, without wavefunctions) and couple it with the translation operator (or the subsidiary condition) to find the wavefunctions; explain why the energy levels are discrete and describe how to use the node theorem to verify completeness.
\item Use quantum principles employed in the class to explain properties of hydrogen including how to use spectroscopy to measure its mass, the radius of the nucleus, how it is used in radio astronomy, and how the probability distribution of electrons are directly measured in momentum space.
\item Use approximation methods of perturbation theory and the variational method to learn about quantum systems that cannot be solved exactly.
\item Calculate how time evolution is employed to describe the oscillatory motion of the simple harmonic oscillator (including coherent and squeezed states) and the rotational motion of a nuclear spin; be able to construct and use the Trotter form of the time-ordered product.
\item Quantize light, describe what a multimode photon is, how single photons are measured, how to verify you have a single-photon light source, how to squeeze light, and how quantum properties of light are used in gravitational wave detection.
\end{enumerate}

One way to summarize the difference between our approach to quantum mechanics and more traditional approaches is the emphasis on physics as opposed to math, and especially the emphasis on experiment. As an example, a traditional class will teach the uncertainty principle stressing the notion that one cannot create quantum states that simultaneously have well-defined position and momentum. Often, this notion is then summarized with the incorrect statement that one cannot measure position and momentum at the same time because the projective measurement projects on one or the other eigenstate only. But, the traditional instructional framework never actually tells us how we measure the momentum of a quantum particle. In our approach, where we emphasize experiments, we spend an entire lecture period discussing the time-of-flight experiment, which is the most common way to measure momentum. It works by having a clock start when a specific event releases a quantum particle whose momentum is to be measured. This can be the release of a trapped atom from its confining trap, or the moment a pulse of high-frequency light impinges on the surface of a metal that photoemits an electron. Then, we time how long it takes until we detect the particle in some detector located a specific distance away from the source. By knowing the distance traveled, the time evolved, and the mass of the particle, we can infer its momentum via a position measurement! In fact, most techniques to measure momentum do so by measuring position and inferring momentum in a similar fashion\cite{momentum}. Wouldn't it be a shame for a quantum-mechanics student not to know how one measures the momentum of a quantum particle? And yet the vast majority of quantum students never learn how this is actually done.

We cover far more experiments as well. The course starts with a conceptual unit that covers many different experiments. We begin with the Stern-Gerlach experiment, but quickly move into delayed choice variants and more sophisticated variants that allow us to perform Bell experiments. We also  discuss interaction-free experiments and reversible tagging in this platform. 

Next, we discuss the properties of light in terms of partial reflection followed by single and multiple-slit diffraction. Then we move on to the Mach-Zehnder interferometer, allowing us to emphasize complementarity with a delayed-choice variant that uses perpendicular polarizers at the two slits. We also discuss a sophisticated quantum seeing in the dark experiment and the Hong-Ou-Mandel experiment. All of this is done before we really jump into formal developments.

Students also learn about teleportation, quantum key distribution, nuclear magnetic resonance, precision experiments that allow us to measure the mass and the volume of the nucleus of hydrogen (including how deuterium was discovered). We also discuss the famous Pickering-Fowler lines from astronomy and how Bohr's explanation of them convinced Einstein that Bohr's theory must be correct. 

Further experiments include e-2e spectroscopy to measure the hydrogen wavefunction in momentum space, how the hyperfine structure is used in radio astronomy, how to measure the wavefunction of an atom in a harmonic trap by time-of-flight, and rotational spectra of molecules.

In the end of the class, after we have quantized light, we describe how a photomultiplier tube works to measure individual quanta, how to verify a single-photon source (which is never a dim laser), and how to measure optical photons using homodyning and heterodyning. We end with a description of how the Laser Interferometry Gravitational Wave Observatory works.

Discussing all of these experiments in a one semester class helps us bring physics back into the quantum-mechanics classroom. After all, we should be teaching physics not math!

\section{CONCLUSIONS}

Quantum mechanics textbooks (and their associated courses) have hardly changed since the 1950s. There have been discussions about the order for how the material should be presented (the famous spins-first or wavefunctions-first approaches) but the material used has remained quite similar to materials that appear in the third edition of Dirac's text,\cite{dirac-3rd} or Schiff's first edition text.\cite{schiff} We argue here that the time is ripe to re-evaluate just how we teach quantum mechanics and that we modernize classes to bring in 21st century topics that help prepare students for the second quantum revolution. We also want to broaden the audience of students who can take such classes. We have described how working with a representation-independent formalism allows us to trade-off higher-level mathematics for abstraction. But, for such an approach to be effective, we must be able to lead students who are not familiar with abstraction to be able to think more abstractly. In our course, we accomplish this by introducing abstraction slowly after we have established quantum ideas in a conceptual unit that requires about 30\% of the total available class time. 
By enabling students to work abstractly, we bring the complex quantum world to them with far fewer prerequisites. This then allows for more coverage of experiment including how to measure individual quanta, which brings students closer to modern topics.

The structure of this course allows for students to learn the foundations needed for future study in quantum sensing. Students leave the course not only knowing how to measure the momentum of a particle, but also how quantum mechanics allows us to measure gravity waves,
What would follow-on classes be---quantum computation and communication is one class that can follow this one. The students are well-prepared to learn how to move into both of those realms (although they would certainly benefit from studying some linear algebra before doing so). But, they also are able to look into how quantum algorithms are used to simulate quantum systems, such as molecules in chemistry, or strongly correlated electrons in condensed-matter physics. They also are well positioned to move onto hands-on classes that work with the experimental equaipment used in quantum sensing. By focusing on modern quantum ideas, this class becomes a gateway to all of quantum information science.

\acknowledgments 
 
We acknowledge useful discussions with Beth Lindsay and Justyna Zwolak. This material is based upon work supported by the Air Force Office of Scientific Research under award
number FA9550-22-1-0469. J.K.F.~was also supported by the McDevitt bequest at Georgetown University.


\end{document}